\documentclass[prb, twocolumn,superscriptaddress,preprintnumbers,amsmath,amssymb, showpacs]{revtex4}

\usepackage{graphicx}
\usepackage{latexsym}
\usepackage{dcolumn}
\usepackage{bm}

\usepackage{float}
\usepackage{graphicx,here}
\usepackage{ulem}
\usepackage{color}

\begin{document}

\title{Interface electronic structure at the topological insulator - ferrimagnetic insulator junction}

\author{Y. Kubota}
\affiliation{Institute for Solid State Physics, The University of Tokyo, Kashiwa, Chiba 277-8581, Japan}
\author{K. Murata}
\affiliation{Department of Electrical Engineering, University of California, Los Angeles, California 90095, United States}
\author{J. Miyawaki}
\affiliation{Institute for Solid State Physics, The University of Tokyo, Kashiwa, Chiba 277-8581, Japan}
\author{K. Ozawa}
\affiliation{Department of Chemistry and Materials Science, Tokyo Institute of Technology, Meguro, Tokyo 152-8551, Japan}
\author{M. C. Onbasli}
\affiliation{Department of Materials Science and Engineering, Massachusetts Institute of Technology, Cambridge, Massachusetts 02139, United States}
\author{T. Shirasawa}
\affiliation{Institute for Solid State Physics, The University of Tokyo, Kashiwa, Chiba 277-8581, Japan}
\author{B. Feng}
\affiliation{Institute for Solid State Physics, The University of Tokyo, Kashiwa, Chiba 277-8581, Japan}
\author{Sh. Yamamoto}
\affiliation{Institute for Solid State Physics, The University of Tokyo, Kashiwa, Chiba 277-8581, Japan}
\author{R.-Y. Liu}
\affiliation{Institute for Solid State Physics, The University of Tokyo, Kashiwa, Chiba 277-8581, Japan}
\author{S. Yamamoto}
\affiliation{Institute for Solid State Physics, The University of Tokyo, Kashiwa, Chiba 277-8581, Japan}
\author{S. K. Mahatha}
\affiliation{Istituto di Struttura della Materia, Consiglio Nazionale delle Ricerche, Trieste, Italy}
\author{P. Sheverdyaeva}
\affiliation{Istituto di Struttura della Materia, Consiglio Nazionale delle Ricerche, Trieste, Italy}
\author{P. Moras}
\affiliation{Istituto di Struttura della Materia, Consiglio Nazionale delle Ricerche, Trieste, Italy}
\author{C. A. Ross}
\affiliation{Department of Materials Science and Engineering, Massachusetts Institute of Technology, Cambridge, Massachusetts 02139, United States}
\author{S. Suga}
\affiliation{Institute of Scientific and Industrial Research, Osaka University, Ibaraki, Osaka 567-0047, Japan}
\author{Y. Harada}
\affiliation{Institute for Solid State Physics, The University of Tokyo, Kashiwa, Chiba 277-8581, Japan}
\author{K. L. Wang}
\affiliation{Department of Electrical Engineering, University of California, Los Angeles, California 90095, United States}
\author{I. Matsuda}
\email{imatsuda@issp.u-tokyo.ac.jp}
\affiliation{Institute for Solid State Physics, The University of Tokyo, Kashiwa, Chiba 277-8581, Japan}

\date{\today}

\begin{abstract}
An interface electron state at the junction between a three-dimensional topological insulator (TI) film of Bi$_2$Se$_3$ and a ferrimagnetic insulator film of Y$_3$Fe$_5$O$_{12}$ (YIG) was investigated by measurements of angle-resolved photoelectron spectroscopy and X-ray absorption magnetic circular dichroism (XMCD).
The surface state of the Bi$_2$Se$_3$ film was directly observed and localized 3$d$ spin states of the Fe$^{3+}$ state in the YIG film were confirmed.
The proximity effect is likely described in terms of the exchange interaction between the localized Fe 3$d$ electrons in the YIG film and delocalized electrons of the surface and bulk states in the Bi$_2$Se$_3$ film.
The Curie temperature ($T_{\mathrm{C}}$) may be increased by reducing the amount of the interface Fe$^{2+}$ ions with opposite spin direction observable as a pre-edge in the XMCD spectra.
\end{abstract}


\maketitle

\section{Introduction}
Topological insulators (TIs) are notable materials currently attracting a wide interest in both fundamental and applied research\cite{Qi, Moore, Hasan}.
Although TIs show bulk insulating performance, they exhibit Dirac-like gapless bands at their surfaces\cite{Xia, HZhang, Hsieh, Sakamoto, Zhang, Bahramy, Landolt}.
The surface state is ensured by time-reversal symmetry (TRS) and the spin polarization of the  surface state electrons is locked to its momentum.
Because these properties are resistant to non-magnetic external perturbations, TIs are expected to be promising materials for new spintronic devices\cite{Qi, Moore, Hasan}.

By breaking TRS, TIs exhibit a number of interesting features, such as the gap-opening at the Dirac point\cite{Chen, Wray}, the half quantum Hall effect\cite{Nomura}, the quantum anomalous Hall effect \cite{Chang, Kou}, the topological magnetoelectric effect\cite{Nomura, Qi_2} and the image magnetic monopole effect\cite{Qi_3}.
There are two methods for breaking the TRS; one is by doping magnetic impurities (Cr, Fe and Mn)\cite{Hor, Chang, Chen, Checkelsky, Kou}, and the other is by connecting TIs to magnetic materials (such as Fe, Co and EuS)\cite{Wray, LiJ, West, Wei, Yang}.
However, with the objective of device applications, magnetic metals in contact with TIs are not appropriate because the TI surface state is short circuited by the metallic material\cite{Liu}.
Furthermore, the Curie temperatures ($T_{\mathrm{C}}$) of these TIs that are doped by magnetic impurities or contacted with the magnetic insulator (EuS) are typically below $35$~K and, therefore, raising $T_{\mathrm{C}}$ is required for practical applications.

Recently, it was suggested that a ferrimagnetic insulator, yttrium iron garnet (YIG, Y$_3$Fe$_5$O$_{12}$) with $T_{\mathrm{C}}$ $\sim$ $550$~K, has the potential to be an underlayer for magnetic TI films\cite{Lang, Liu, Jiang}.
It was reported, from magneto-transport and magneto-optical measurements, that $T_{\mathrm{C}}$ of the Bi$_2$Se$_3$/YIG system reaches $\sim 130$~K\ due to the proximity effect\cite{Lang}.
Furthermore, $T_{\mathrm{C}}$ of Cr-doped Bi$_2$Se$_3$ on YIG was found to be higher than that on a nonmagnetic substrate through magneto-transport and X-ray absorption magnetic circular dichroism (XMCD) measurements\cite{Liu}.
Understanding the mechanism of the proximity effect between TI and YIG is required to realize $T_{\mathrm{C}}$ above room temperature (RT).

In this paper, we present results of angle-resolved photoelectron spectroscopy (ARPES) and XMCD for Bi$_2$Se$_3$ films on YIG.
We have successfully observed the TI surface state in this Bi$_2$Se$_3$/YIG system and obtained direct evidence that the $3d$ electrons of Fe in YIG induce the proximity effect at the interface between TI and YIG.
This result represents a potential advance for the optimization of the properties of the magnetic layer to realize practical new devices.

\begin{figure}[htb]
\includegraphics[width=8cm]{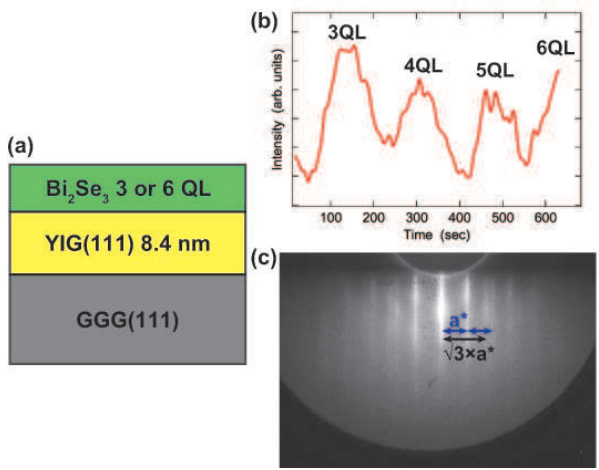}
\caption{(a) A schematic drawing of the Bi$_2$Se$_3$ film prepared on the YIG$\left(111\right)$/GGG$\left(111\right)$ substrate. (b) The $\left( 00 \right)$-spot RHEED intensity oscillation during Bi$_2$Se$_3$ growth on the YIG film, taken at an electron energy of $15$~keV.
(c) The RHEED pattern of a $6$~QL Bi$_2$Se$_3$/YIG sample.
The symbol a$^\ast$ represents the reciprocal lattice constant of Bi$_2$Se$_3$.}
\label{sample}
\end{figure}

\section{Experiment}
Figure \ref{sample} (a) shows a schematic drawing of the sample.
YIG (111) thin films ($8.4$~nm thick) were prepared on gadolinium gallium garnet (GGG) (111) substrates\cite{Lang, Liu}.
The details of YIG growth can be found in the Supplemental Material\cite{SI}.
Bi$_2$Se$_3$ films were then grown on YIG/GGG using Bi and Se conventional effusion cells.
At first, 2 quintuple layers (QLs) Bi$_2$Se$_3$ films were grown on YIG which was maintained at $150^\circ$C. The samples were then annealed at $300^\circ$C, followed by further Bi and Se deposition at $250^\circ$C.
The thickness of the Bi$_2$Se$_3$ films was controlled from $3$ to $6$~QL by observation of reflection high-energy electron diffraction (RHEED) intensity oscillations, as shown in Fig. \ref{sample} (b).
After the deposition of the Bi$_2$Se$_3$ films at $250^\circ$C, the samples were further annealed at $250^\circ$C for 15 min to realize a better crystalline quality.
Figure \ref{sample} (c) presents the RHEED pattern of the $6$~QL Bi$_2$Se$_3$/YIG sample. Referred from the $\left( 00 \right)$-rod, streaks are identified at $2$a$^\ast$ and $\sqrt{3} \times $a$^\ast$ where a$^\ast$ represents the reciprocal lattice constant of Bi$_2$Se$_3$. The pattern indicates that the Bi$_2$Se$_3$ films have a multi-domain structure such as a $\left < 111 \right>$-oriented texture structure\cite{Gotoh}. The sample surface was capped with a $30$~nm-thick Se layer and the sample wafer was transferred in air to the ARPES measurement chamber.

ARPES measurements for the $6$~QL Bi$_2$Se$_3$ film were performed on the VUV-Photoemission beamline at Elettra.
After removing the Se capping layer by annealing at $190^\circ$C, {\it p}-polarized light was incident onto the sample.
Measurements of XMCD were performed on $3$ and $6$~QL Bi$_2$Se$_3$ films on the YIG films at RT and $20$~K at the BL07LSU beamline\cite{Yamamoto} at SPring-8.

\begin{figure*}[htb]
\includegraphics[width=15cm]{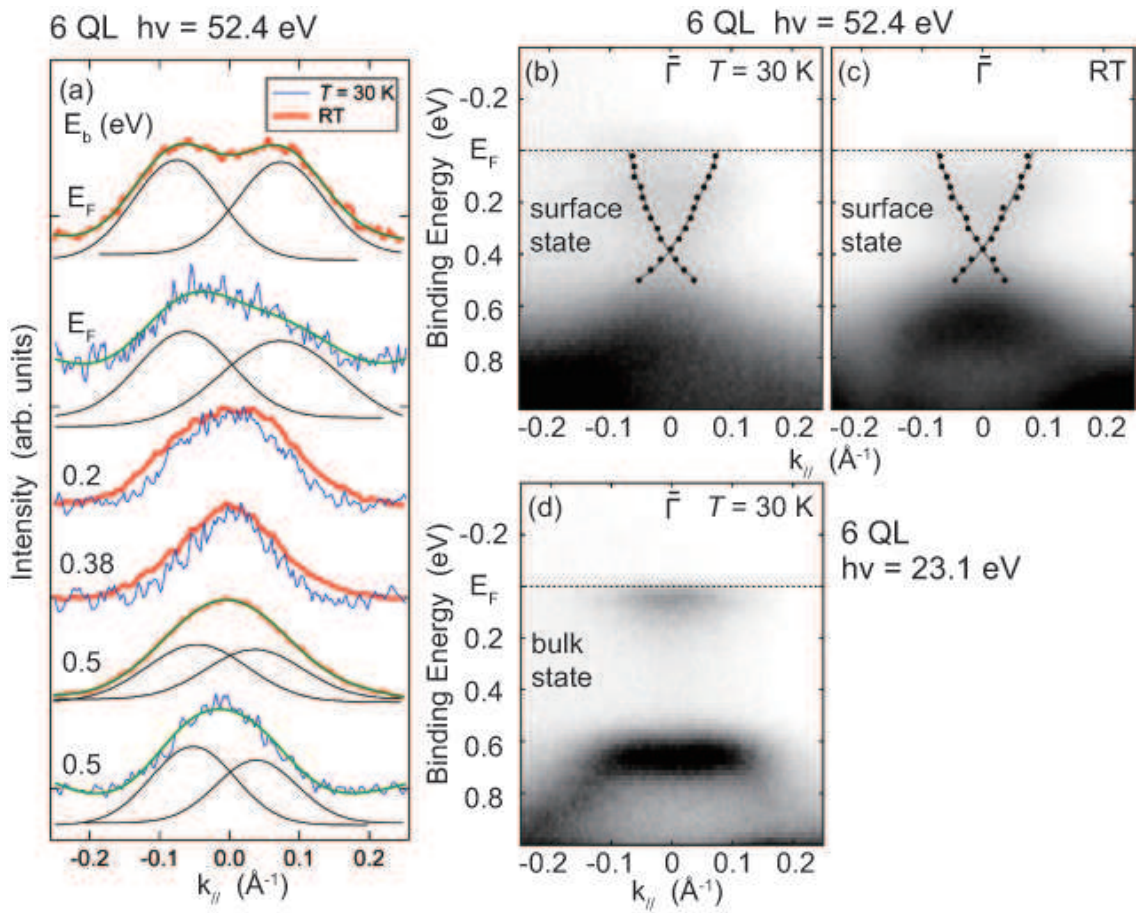}
\caption{(a) Photoelectron momentum distribution curves (MDCs), taken at h$\rm \nu = 52.4$~eV, for the $6$~QL Bi$_2$Se$_3$/YIG sample at $T = 30$~K (blue solid lines) and RT (red solid lines). The solid black and green lines indicate the two-peak fitting curves. (b, c) Photoelectron band diagrams around $\Bar{\Gamma}$ for the $6$~QL Bi$_2$Se$_3$/YIG sample at (b) $T = 30$~K and (c) RT (h$\rm \nu = 52.4$~eV).The solid circles correspond to the peak position from the MDCs and the black lines are fits.(d) ARPES spectra of the $6$~QL Bi$_2$Se$_3$/YIG sample at $T = 30$~K (h$\rm \nu = 23.1$~eV).}
\label{ARPES}
\end{figure*}

\section{Result and discussion}
\subsection{Angle-resolved photoelectron spectroscopy}
Figure \ref{ARPES} (a) shows the momentum ($k_{//}$) distribution curves (MDCs) of the ARPES spectra around the $\Bar{\Gamma}$ point, taken at h$\rm \nu=52.4$~eV, at $T = 30$~K and RT.
The MDCs obtained at $T = 30$~K have slightly narrower peaks than those at RT due to reduction of the thermal blurring.
In the figure, it can be seen that the two peaks in the MDCs at the Fermi level ($E_{\mathrm{F}}$) approach with binding energy ($E_{\mathrm{b}}$) and overlap each other at $E_{\mathrm{b}} = 0.38$~eV, followed by separation at higher $E_{\mathrm{b}}$.
These results unambiguously indicate the  band-crossing.
The MDC peaks in Fig. \ref{ARPES} (a) were curve-fitted by two peaks and the peak positions are plotted in the photoelectron band diagram in Fig. \ref{ARPES} (b, c).
At $T = 30$~K and RT, the surface state band shows the Dirac-cone dispersion around the $\Bar{\Gamma}$ point with the Dirac point at $E_{\mathrm{b}} = 0.38$~eV.
The band-dispersion curves were assigned to those of the Dirac surface state bands of the Bi$_2$Se$_3$ film as reported previously\cite{Xia, Landolt}.
The Fermi velocity of this system is $v_{\mathrm{F}} = 5.1 \times 10^5$~m/s and this value agrees with previous studies\cite{HZhang, Sakamoto, Zhang}.
For comparison, the photoelectron band diagram, taken at h$\rm \nu = 23.1$~eV, is also shown in Fig. \ref{ARPES} (d).
The observed band between $E_{\mathrm{F}}$ and $0.2$~eV is assigned to the bulk conduction band\cite{Xia, Landolt} of Bi$_2$Se$_3$  and it crosses the $E_{\mathrm{F}}$, indicating the {\it n}-type doped nature.
The electronic structures of the surface and bulk states are essentially similar to the previous ARPES results of Bi$_2$Se$_3$ films on different substrates\cite{Xia, Hsieh, Sakamoto, Zhang, Bahramy, Landolt}.
Due to the TI nature of Bi$_2$Se$_3$ film\cite{Qi, Moore, Hasan}, the existence of the surface state at the film/vacuum interface suggests its presence also at the junction (interface) with the YIG film.
Moreover, Fig. \ref{ARPES} (d) implies that the Bi$_2$Se$_3$ bulk conduction band crosses $E_{\mathrm{F}}$ at the Bi$_2$Se$_3$/YIG interface.
It is of note that the surface state band structure does not change with the temperature across $T_{\mathrm{C}} \sim 130$~K\cite{Lang}, as shown in Fig. \ref{ARPES}.

\begin{figure*}[htb]
\includegraphics[width=15cm]{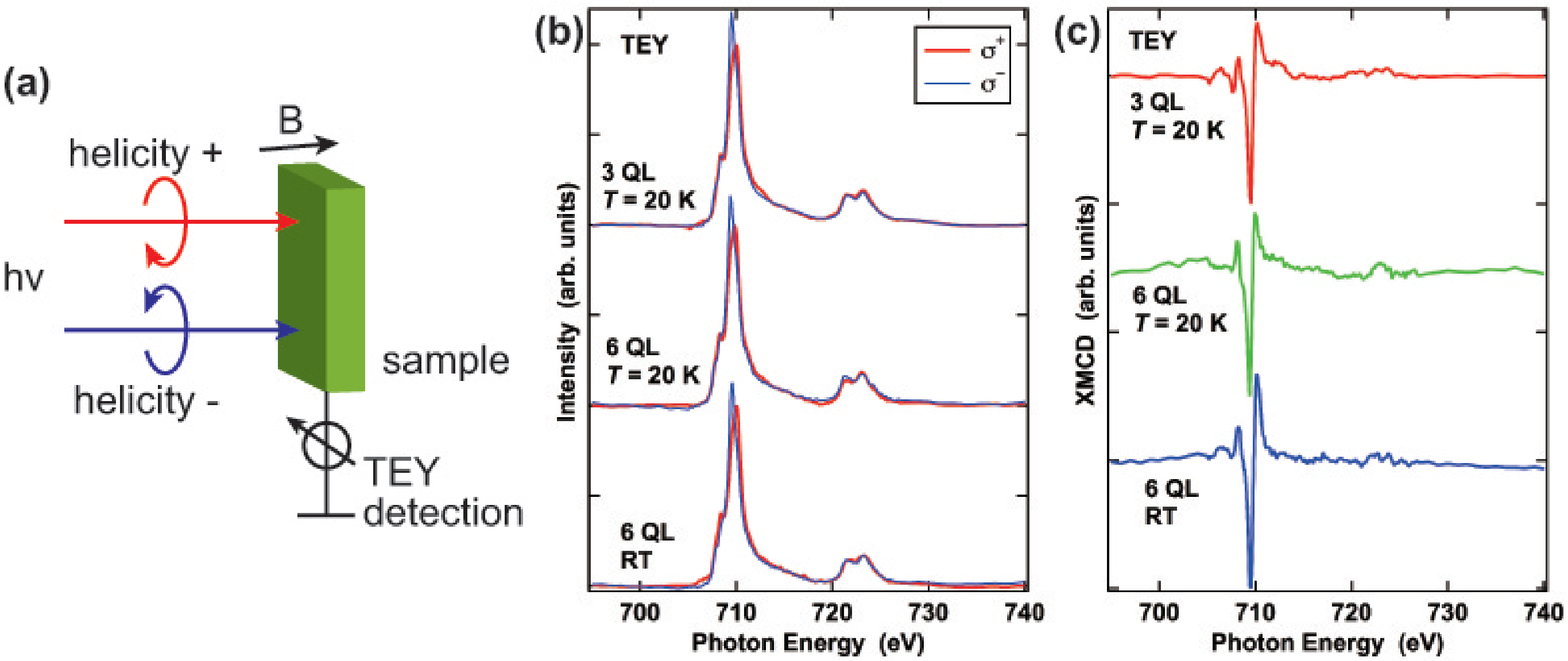}
\caption{(a) A schematic drawing of the XMCD experimental set up.
(b) XAS spectra of the $3$ and $6$~QL Bi$_2$Se$_3$/YIG samples obtained by the total electron yield (TEY) detection at RT and $20$~K. The solid red and blue lines represent the spectra taken with circular-polarized light of plus and minus helicity ($\sigma^+$, $\sigma^-$), respectively.
(c) XMCD spectra  of $3$~QL Bi$_2$Se$_3$/YIG at $T = 20$~K (red), $6$~QL Bi$_2$Se$_3$/YIG at $T = 20$~K (green), and $6$~QL Bi$_2$Se$_3$/YIG at RT (blue).}
\label{MCD}
\end{figure*}

\subsection{X-ray absorption magnetic circular dichroism}
Figure \ref{MCD} (a) shows the XMCD measurement configuration.
A magnetic field of 0.24 T was applied by a retractable permanent magnet.
The XMCD was measured at the Fe $L_{2,3}$-shell absorption edge of YIG by the total electron yield (TEY) mode.
XMCD spectra were derived from the difference between the two adsorption spectra obtained by circularly polarized light of opposite helicities, where the beam direction was set parallel to the magnetic field orientation and to the surface normal direction.
Figures \ref{MCD} (b, c) show the Fe ${2p}$ X-ray absorption spectra (b) and XMCD (c).
The spectral shapes are mostly in agreement with those of the Cr-doped Bi$_2$Se$_3$/YIG sample reported by Liu {\it et al.}\cite{Liu}.
Since the probing depth of the present XMCD measurement is about $5 \sim 10$~nm not much different from the thickness of $3$~nm ($3$~QL) and $6$~nm ($6$~QL) Bi$_2$Se$_3$\cite{Zhao, Stohr}, the XMCD signals are thought to be essentially resulting from the Fe atoms near the Bi$_2$Se$_3$/YIG boundary.
The positive and negative XMCD peaks at the $L_3$-edge suggest the opposite spin direction in the Fe atoms, at two different sites in the YIG crystal, octahedral (2 per formula unit) and tetrahedral sites (3 per formula unit), as expected for the ferrimagnet.
In the present measurement configuration, the macroscopic magnetic direction of the ferrimagnetic YIG film shows a negative peak for the tetrahedral Fe site \cite{Liu}.
While the overall Fe ${2p}$ X-ray absorption spectra and XMCD are similar in spite of the different thickness and temperature, we notice a slight difference at the pre-edge structure that may depend on thickness and temperature as described below.

\begin{figure}[htb]
\includegraphics[width=8cm]{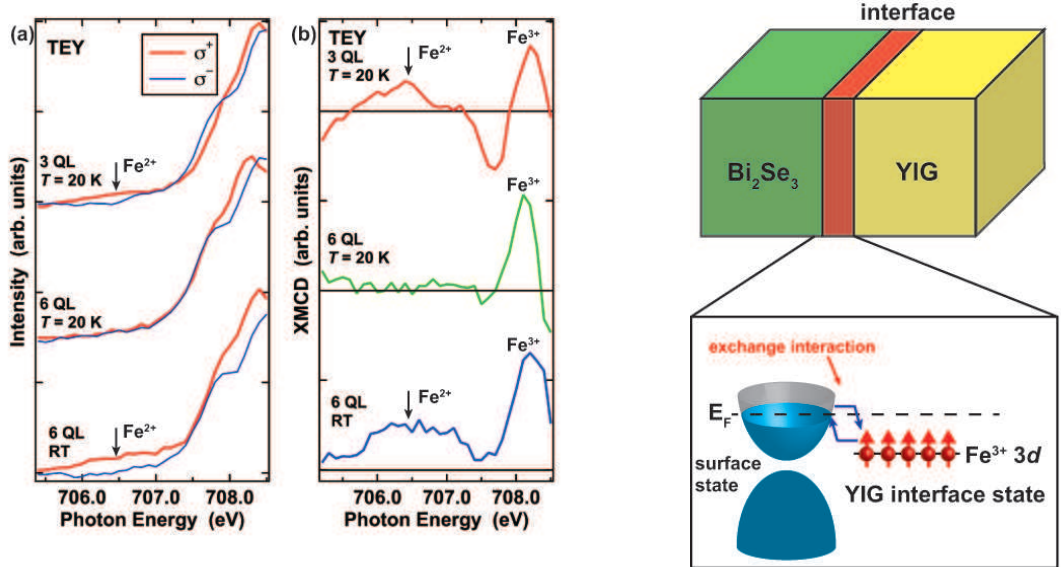}
\caption{Enlarged figures of the (a) XAS and (b) XMCD spectra in Fig. \ref{MCD}. The solid horizontal lines in (b) represent $\mathrm{XMCD} = \mathrm{XAS} \left( \sigma^+ \right) - \mathrm{XAS} \left( \sigma^- \right) = 0$.
Peaks of the $\sigma^+$ and XMCD spectra in this photon energy region are observed in the data of the $3$~QL Bi$_2$Se$_3$/YIG at $T = 20$~K and $6$~QL Bi$_2$Se$_3$/YIG at RT, as shown by arrows.}
\label{MCD_close}
\end{figure}

\begin{figure}[htb]
\includegraphics[width=8cm]{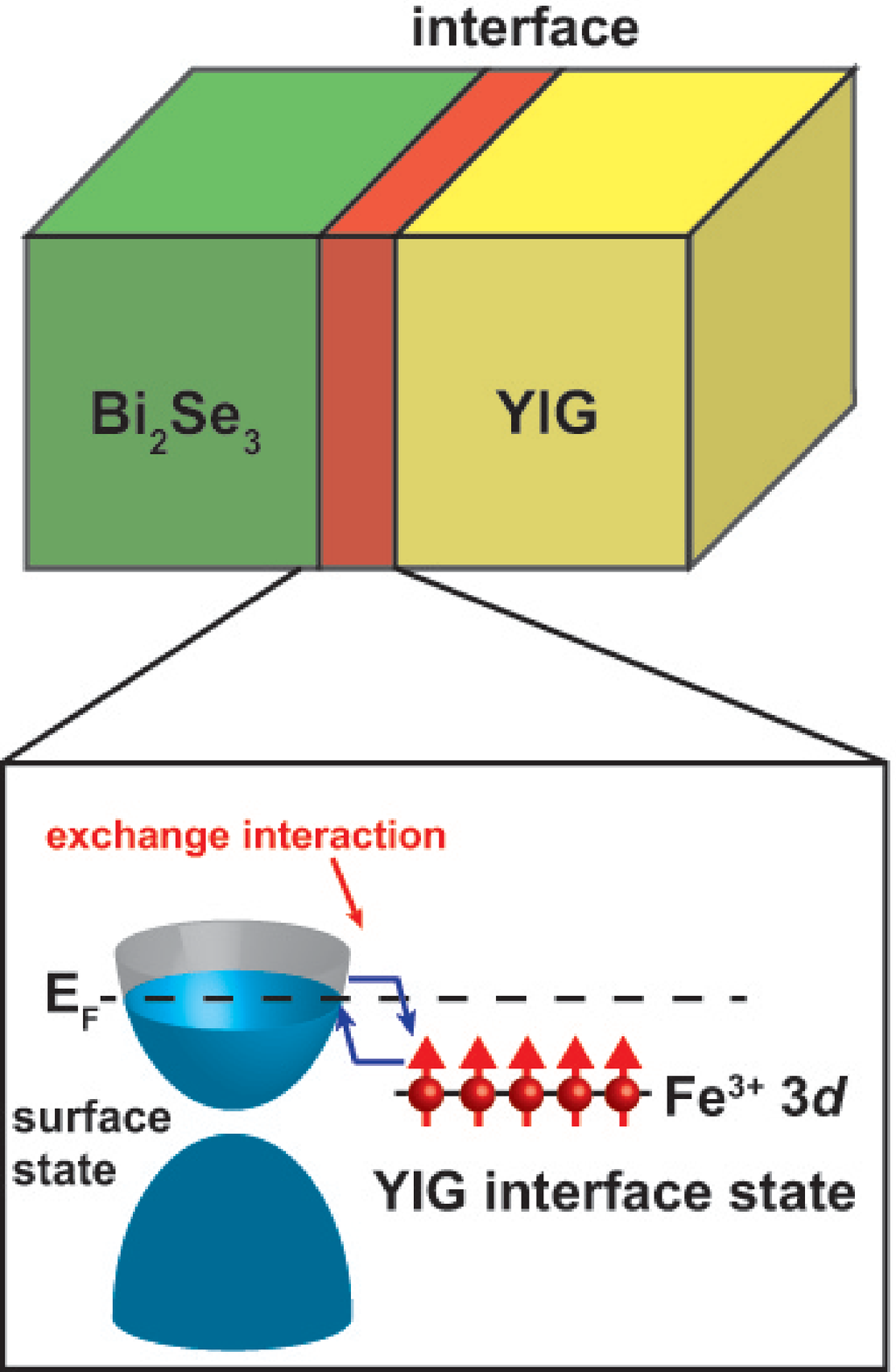}
\caption{A schematic drawing of the proximity effect at the Bi$_2$Se$_3$ and YIG interface. The origin is the exchange interaction between the spin-polarized electrons of the Bi$_2$Se$_3$ film and the localized $3d$ electrons of the Fe$^{3+}$ sites.The Bi$_2$Se$_3$ surface state band produces gap-opening at the Dirac point. }
\label{model}
\end{figure}

Figure \ref{MCD_close} shows enlarged images of the X-ray absorption (a) and XMCD (b) spectra in h$\rm \nu = 705.2 \sim 708.5$~eV.
One can notice peak shoulders at h$\rm \nu \sim 706.5$~eV in the $\sigma^+$ spectra of the $3$~QL Bi$_2$Se$_3$ film at $T = 20$~K and the $6$~QL Bi$_2$Se$_3$ film at RT.
Compared with the previous X-ray absorption study\cite{Miedema}, the spectral features are likely assigned to the Fe$^{2+}$ state in the YIG crystal.
These results indicate the existence of the Fe$^{2+}$ state at the Bi$_2$Se$_3$/YIG interface.
The pre-edge peak (h$\rm \nu \sim 706.5$~eV) shows a positive XMCD signal and, thus, it is natural to assume that the Fe$^{2+}$ possesses an opposite spin direction from the net magnetization of the YIG crystal.
It is intriguing to note that the pre-edge feature is seemingly suppressed in the X-ray absorption and XMCD spectra of the $6$~QL Bi$_2$Se$_3$ film at $T = 20$~K.
We infer that these Fe$^{2+}$ states with the opposite spin direction may prevent the ferromagnetism of Bi$_2$Se$_3$ at the interface.
While the interface Fe$^{2+}$ state is considered to be related to the Bi-Fe or Se-Fe interaction at the Bi$_2$Se$_3$/YIG junction, the proper assignment requires determination of the accurate interface atomic structure and appropriate theoretical calculations.

\subsection{Spin interaction at the interface}
From the experimental results above, ferromagnetism of Bi$_2$Se$_3$ at the interface would be associated with the interface spin-polarized states of the Bi$_2$Se$_3$ film and localized spin states of the interface Fe$^{3+}$ in the YIG film.
Thus, the proximity effect would be modeled as their spin exchange interactions at the boundary, as shown in Fig. \ref{model}.
Such interface interaction has been already calculated for a similar system that is composed of a Bi$_2$Se$_3$ film and an EuS substrate \cite{Li}.
When the TI film has a surface state with a gap at the Dirac point by breaking the TRS, these delocalized spins were found to experience an exchange interaction with the localized spins of the $4f$ electrons in the Eu$^{2+}$ ions\cite{Li}.
Moreover, it was found that bulk ($p_{\mathrm{z}}$-orbital) states of the Bi$_2$Se$_3$ film also contribute to the spin-coupling between the TI and the ferromagnetic material when $E_{\mathrm{F}}$ locates above the minimum of the bulk conduction band\cite{Li}.
By analogy, the ferromagnetism of Bi$_2$Se$_3$ at the interface can be understood as an exchange interaction between the spin-polarized electrons in the (gapped) Dirac surface state of the Bi$_2$Se$_3$ film and the localized $3d$ electrons of the Fe$^{3+}$ in the YIG film\cite{Lang, Wei}.
Moreover, there is also a contribution from the bulk electrons in the Bi$_2$Se$_3$ film since the bulk conduction band crosses $E_{\mathrm{F}}$, as shown in Fig. \ref{ARPES} (d).

\section{Summary}
In summary, in the present research, we provide the first evidence of the surface state of the Bi$_2$Se$_3$ film on YIG by ARPES and the significance of the Fe$^{3+}$ state for ferromagnetism of the Bi$_2$Se$_3$ at the interface by Fe $L_{2, 3}$-edge XMCD.
The origin of the proximity effect is likely described in terms of the exchange interaction between the localized Fe$^{3+}$ 3$d$ electrons in the YIG film and the delocalized electrons of the surface state and the bulk state in the Bi$_2$Se$_3$ film.
One may increase $T_{\mathrm{C}}$ for developing future spintronic devices by reducing the amount of Fe$^{2+}$ ions with opposite spin direction that may exist at the interface.

\section*{Acknowledgments}
We thank N. Fukui, R. Hobara and S. Hasegawa for their assistance in the sample preparation, and gratefully acknowledge A. Kimura for valuable discussion.
Y. Kubota acknowledges support from the ALPS programs of the University of Tokyo.
This work was partially supported by the Ministry of Education, Culture, Sports, Science, and Technology of Japan (X-ray Free Electron Laser Priority Strategy Program and Photon and Quantum Basic Research Coordinated Development Program) and performed using facilities of the Synchrotron Radiation Research Organization, The University of Tokyo (Proposal No. 2014B7473, 2015B7401, 2015A7401, 2014B7401, 2014A7401).
UCLA work was partially supported from the Spins and Heat in Nanoscale Electronic Systems (SHINES), an Energy Frontier Research Center funded by the U.S. Department of Energy (DOE), Office of Science, Basic Energy Sciences (BES), under Award No. DE-SC0012670, and from the U.S. Army Research Office under award $\#$ W911NF-15-1-0561.
C. A. Ross and M. C. Onbasli are very grateful to the support from the FAME Center, one of six centers of STARnet, a Semiconductor Research Corporation program sponsored by MARCO and DARPA.
P. Moras acknowledges financial support from CNR in the framework of the agreement between CNR and JSPS (Japan), project "Study of spin and electronic properties of topological insulator films regulated by the surface and the interface atomic layers".


\begin{thebibliography}{99}
{
\bibitem{Qi} X.-L. Qi and S.-C. Zhang, Rev. Mod. Phys. {\bf 83}, 1057 (2011).
\bibitem{Moore} J. E. Moore, Nature {\bf 464}, 194 (2010).
\bibitem{Hasan} M. Z. Hasan and C. L. Kane, Rev. Mod. Phys. {\bf 82}, 3045 (2010).
\bibitem{Xia} Y. Xia, D. Qian, D. Hsieh, L. Wray, A. Pal, H. Lin, A. Bansil, D. Grauer, Y. S. Hor, R. J. Cava, and M. Z. Hasan, Nat. Phys. {\bf 5}, 398 (2009).
\bibitem{HZhang} H. Zhang, C.-X. Liu, X.-L. Qi, X. Dai, Z. Fang, and S.-C. Zhang, Nat. Phys. {\bf 5}, 438 (2009).
\bibitem{Hsieh} D. Hsieh, Y. Xia, D. Qian, L. Wray, J. H. Dil, F. Meier, J. Osterwalder, L. Patthey, J. G. Checkelsky, N. P. Ong, A. V. Fedorov, H. Lin, A. Bansil, D. Grauer, Y. S. Hor, R. J. Cava, and M. Z. Hasan, Nature {\bf 460}, 1101 (2009).
\bibitem{Sakamoto} Y. Sakamoto, T. Hirahara, H. Miyazaki, S.I. Kimura, and S. Hasegawa, Phys. Rev. B {\bf 81}, 165432 (2010).
\bibitem{Zhang} Y. Zhang, K. He, C.-Z. Chang, C.-L. Song, L.-L. Wang, X. Chen, J.-F. Jia, Z. Fang, X. Dai, W.-Y. Shan, S.-Q. Shen, Q. Niu, X.-L. Qi, S.-C. Zhang, X.-C. Ma, and Q.-K. Xue, Nat. Phys. {\bf 6}, 584 (2010).
\bibitem{Bahramy} M. S. Bahramy, P. D. C. King, A. de la Torre, J. Chang, M. Shi, L. Patthey, G. Balakrishnan, P. Hofmann, R. Arita, N. Nagaosa, and F. Baumberger, Nat. Commun. {\bf 3}, 1159 (2012).
\bibitem{Landolt} G. Landolt, S. Schreyeck, S. V. Eremeev, B. Slomski, S. Muff, J. Osterwalder, E. V. Chulkov, C. Gould, G. Karczewski, K. Brunner, H. Buhmann, L. W. Molenkamp, and J. H. Dil, Phys. Rev. Lett. {\bf 112}, 057601 (2014).
\bibitem{Chen} Y. L. Chen, J.-H. Chu, J. G. Analytis, Z. K. Liu, K. Igarashi, H.-H. Kuo, X. L. Qi, S. K. Mo, R. G. Moore, D. H. Lu, M. Hashimoto, T. Sasagawa, S. C. Zhang, I. R. Fisher, Z. Hussain, and Z. X. Shen, Science {\bf 329}, 659 (2010).
\bibitem{Wray} L. A. Wray, S.-Y. Xu, Y. Xia, D. Hsieh, A. V. Fedorov, Y. S. Hor, R. J. Cava, A. Bansil, H. Lin, and M. Z. Hasan, Nat. Phys. {\bf 7}, 32 (2010).
\bibitem{Nomura} K. Nomura and N. Nagaosa, Phys. Rev. Lett. {\bf 106}, 166802 (2011).
\bibitem{Chang} C.-Z. Chang, J. Zhang, X. Feng, J. Shen, Z. Zhang, M. Guo, K. Li, Y. Ou, P. Wei, L.-L. Wang, Z.-Q. Ji, Y. Feng, S. Ji, X. Chen, J. Jia, X. Dai, Z. Fang, S.-C. Zhang, K. He, Y. Wang, L. Lu, X.-C. Ma, and Q.-K. Xue, Science {\bf 340}, 167 (2013).
\bibitem{Kou} X. Kou, S.-T. Guo, Y. Fan, L. Pan, M. Lang, Y. Jiang, Q. Shao, T. Nie, K. Murata, J. Tang, Y. Wang, L. He, T.-K. Lee, W.-L. Lee, and K. L. Wang, Phys. Rev. Lett. {\bf 113}, 137201 (2014).
\bibitem{Qi_2} X.-L. Qi, T. L. Hughes, and S.-C. Zhang, Phys. Rev. B {\bf 78}, 195424 (2008).
\bibitem{Qi_3} X.-L. Qi, R. Li, J. Zang, and S.-C. Zhang, Science {\bf 323}, 1184 (2009).
\bibitem{Hor} Y. S. Hor, P. Roushan, H. Beidenkopf, J. Seo, D. Qu, J. G. Checkelsky, L. A. Wray, D. Hsieh, Y. Xia, S.-Y. Xu, D. Qian, M. Z. Hasan, N. P. Ong, A. Yazdani, and R. J. Cava, Phys. Rev. B {\bf 81}, 195203 (2010).
\bibitem{Checkelsky} J. G. Checkelsky, J. Ye, Y. Onose, Y. Iwasa, and Y. Tokura, Nat. Phys. {\bf 8}, 729 (2012).
\bibitem{LiJ} J. Li, Z. Y. Wang, A. Tan, P.-A. Glans, E. Arenholz, C. Hwang, J. Shi, and Z. Q. Qiu, Phys. Rev. B {\bf 86}, 054430 (2012).
\bibitem{West} D. West, Y. Y. Sun, S. B. Zhang, T. Zhang, X. Ma, P. Cheng, Y. Y. Zhang, X. Chen, J. F. Jia, and Q. K. Xue, Phys. Rev. B {\bf 85}, 081305 (2012).
\bibitem{Wei} P. Wei, F. Katmis, B. A. Assaf, H. Steinberg, P. Jarillo-Herrero, D. Heiman, and J. S. Moodera, Phys. Rev. Lett. {\bf 110}, 186807 (2013).
\bibitem{Yang} Q. I. Yang, M. Dolev, L. Zhang, J. Zhao, A. D. Fried, E. Schemm, M. Liu, A. Palevski, A. F. Marshall, S. H. Risbud, and A. Kapitulnik, Phys. Rev. B {\bf 88}, 081407 (2013).
\bibitem{Liu} W. Liu, L. He, Y. Xu, K. Murata, M. C. Onbasli, M. Lang, N. J. Maltby, S. Li, X. Wang, C. A. Ross, P. Bencok, G. van der Laan, R. Zhang, and K. L. Wang, Nano Lett. {\bf 15}, 764 (2014).
\bibitem{Lang} M. Lang, M. Montazeri, M. C. Onbasli, X. Kou, Y. Fan, P. Upadhyaya, K. Yao, F. Liu, Y. Jiang, W. Jiang, K. L. Wong, G. Yu, J. Tang, T. Nie, L. He, R. N. Schwartz, Y. Wang, C. A. Ross, and K. L. Wang, Nano Lett. {\bf 14}, 3459 (2014).
\bibitem{Jiang} Z. Jiang, F. Katmis, C. Tang, P. Wei, J. S. Moodera, and J. Shi, Appl. Phys. Lett. {\bf 104}, 222409 (2014).
\bibitem{SI} See Supplemental Material for details of the YIG film preparation.
\bibitem{Gotoh} Y. Gotoh and S. Ino, Thin Solid Films {\bf 109}, 255 (1983).
\bibitem{Yamamoto} S. Yamamoto, Y. Senba, T. Tanaka, H. Ohashi, T. Hirono, H. Kimura, M. Fujisawa, J. Miyawaki, A. Harasawa, T. Seike, S. Takahashi, N. Nariyama, T. Matsushita, M. Takeuchi, T. Ohata, Y. Furukawa, K. Takeshita, S. Goto, Y. Harada, S. Shin, H. Kitamura, A. Kakizaki, M. Oshima, and I. Matsuda, J. Synchrotron Radiat. {\bf 21}, 352 (2014).
\bibitem{Zhao} S. Y. F. Zhao, C. Beekman, L. J. Sandilands, J. E. J. Bashucky, D. Kwok, N. Lee, A. D. LaForge, S. W. Cheong, and K. S. Burch, Appl. Phys. Lett. {\bf 98}, 141911 (2011).
\bibitem{Stohr} J. St$\ddot{\mathrm{o}}$hr, {\it NEXAFS Spectroscopy} (Springer-Verlag, Berlin, 1992), p. 124.
\bibitem{Miedema} P. S. Miedema and F. M. F. de Groot, J. Electron Spectros. Relat. Phenomena {\bf 187}, 32 (2013).
\bibitem{Li} M. Li, W. Cui, J. Yu, Z. Dai, Z. Wang, F. Katmis, W. Guo, and J. Moodera, Phys. Rev. B {\bf 91}, 014427 (2015).

}\end{thebibliography}
\end{document}